\documentclass[a4paper,12pt]{article}

\usepackage{amssymb, amsmath, amsthm}
\usepackage{psfig}
\usepackage{epsfig}
\usepackage{subfigure}   
\usepackage{wrapfig}

\title{Lambda AntiLambda Production as a Benchmark Channel for PANDA}

\author{Marco Destefanis
\\
Justus Liebig Universit\"at Giessen - Germany\\
for the PANDA collaboration}

\begin{document}

\date{}

\maketitle

\begin{abstract}
\label{abstract}
The PANDA \cite{LOI,Technical} experiment which is part of 
the future FAIR facility at
Darmstadt will investigate reactions of antiprotons with hydrogen
and nuclear targets. One of the benchmark channel for the simulation
and the design of the detector is the $\Lambda \bar{\Lambda}$ channel,
which has been extensively investigated by the PS185 collaboration
\cite{PS185web,PS185LOI,PS185-1,PS185-2,PS185-3}. 
This paper will report on first results on GEANT4
simulations including the 
PANDA detector geometry which have been
performed to study the detector acceptance, resolution and background
suppression as well as the reconstruction polarization observables.
\end{abstract}

\section{Physics and Detector}
\label{fisdetPANDA}
The PANDA collaboration \cite{LOI,Technical} proposes to 
build a state-of-the-art universal 
detector for strong interaction studies at the high-energy storage ring 
HESR at the international FAIR facility in Darmstadt. The detector 
is designed to 
take advantage of the extraordinary physics potential which will be
available utilizing high intensity, phase space cooled antiproton beams.

\begin{figure}[h!] 
\begin{center}
\includegraphics[width=0.8\textwidth,keepaspectratio]{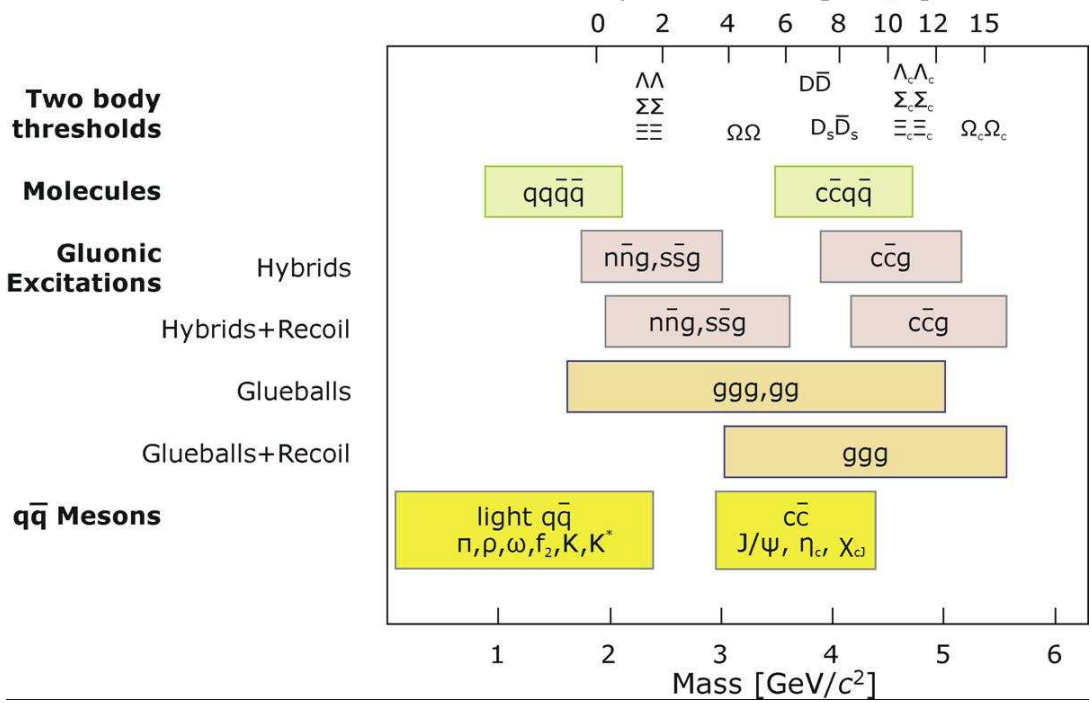}
\caption{\textit{Mass range of hadrons accessible at the HESR with
    antiproton beams.}}
\label{masseaccessibili}
\end{center}
\end{figure}

The collaboration is planning to use the antiproton beams, provided 
by the new FAIR facility, in a momentum range of 1 GeV/c to 15 GeV/c to
address fundamental questions on QCD. The fundamental building blocks
of QCD are the quarks which interact with each other by exchanging
gluons. When the distance between the quarks becomes
comparable to the size of the nucleon, the force among them becomes so
strong that they cannot be further separated, in contrast to the
electromagnetic and gravitational forces which fall off with increasing
distance. This unusal behavior is related to the self-interaction of
gluons: gluons do not only interact with quarks but also with each
other, leading to the formation of gluonic flux tubes connecting the
quarks. As a consequence, quarks have never been observed as free
particles and are confined within hadrons, complex particles made of 3
quarks (baryons) or a quark-antiquark pair (mesons). Baryons and
mesons are the relevant degrees of freedom in our environment. An
important consequence of the gluon self-interaction and a strong proof
of our understanding of hadronic matter is the prediction of
hadronic systems consisting only of gluons (glueballs) or bound
systems of quark-antiquark pairs and gluons (hybrids) with both normal
and exotic quantum numbers. As illustrated in
Fig. \ref{masseaccessibili}, the physics program,
which can be developed using the HESR storage ring, 
offers a broad range of investigation that extend from the
study of QCD to the test of fundamental symmetries.

In antiproton-proton annihilations, particles with gluonic degrees of
freedom as well as particle-antiparticle pairs are copiously
produced, allowing spectroscopic studies with unprecedented statistics
and precision. The following experiments are forseen:
 
\begin{itemize}
\item{}Charmonium ($c\bar{c}$) spectroscopy
\item{}Firm estabilishment of the QCD-predicted gluonic excitations in
  the charmonium mass range (3-5GeV/c$^2$)
\item{}Search for modifications of meson properties in the nuclear
  medium
\item{}Precision$\gamma$-ray spectroscopy of single and double
  hypernuclei
\end{itemize}

Further physics opportunities will open up as soon as the HESR
facility reaches the full design luminosity:

\begin{itemize}
\item{}Extraction of generalized parton distributions from $p\bar{p}$
  annihilations
\item{}$D$ meson decay spectroscopy
\item{}Search for $CP$ violation in the charm and strangeness sector
  ($D$ meson decays, $\Lambda\bar{\Lambda}$ system)
\end{itemize}

The physics program is intensively described in this volume by Tobias
Stockmanns.

The physics program, as described, poses significant challenges for
the PANDA detector. The tasks are summarized for individual detectors:

\begin{itemize} 
\item{}Full angular coverage and good angular resolution for both
  charged and neutral particles
\item{}Particle identification in a large range of particles
  ($\gamma$-rays, e, $\mu$, kaons, protons and pions)
\item{}High resolution for a wide range of energies
\item{}High rate compatibility expecially for the close-to-target and
  forward deterctors
\end{itemize}

\begin{figure}[h!] 
\begin{center}
\includegraphics[width=\textwidth,keepaspectratio]{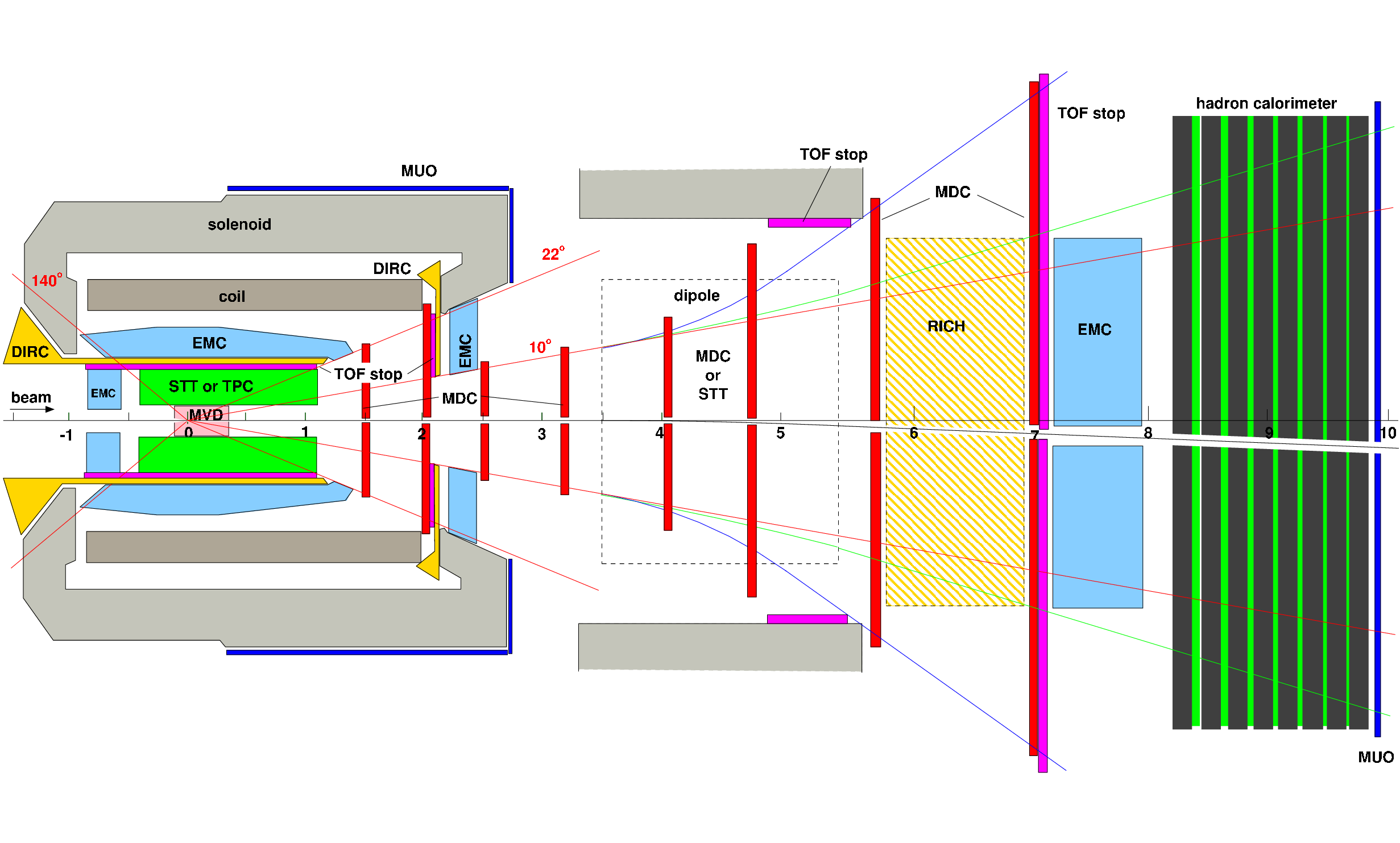}
\caption{\textit{Setup of the PANDA detector.}}
\label{spettrometro}
\end{center}
\end{figure}

The detector consists of two spectrometers. A target spectrometer
(TS) surrounds the interaction region and a forward spectrometer with
a second magnet provides angular coverage for the most forward
angles. The basic concept of the target spectrometer is a shell-like
arrangement of various detector system surrounding the interaction
point inside the field of a large solenoid. The forward spectrometer
will cover the gap in detector acceptance in the forward
region. The spectrometer apparatus is shown in
Fig. \ref{spettrometro}.

The target spectrometer (TS) is an azimuthally symmetric system
of detectors mostly contained inside the superconducting
solenoid. This part includes the muon counters on the outside of the
magnet return yoke and the forward end cap which may be situated just
behind the magnet. This part of PANDA spectrometer will detect all particles
emitted with laboratory angles greater than 5$^\circ$ and 10$^\circ$,
in vertical and horizontal planes respectively
, and lower than 140$^\circ$.
No particle detection is foreseen in very backward region of 170-180
degrees. 
Surrounding the interaction volume there will be a silicon
micro-vertex detector (MVD). A second tracking detector will
consist of 15 double layers of straw tubes (STT). 
Another possiblity  is to use time projection chamber (TPC) instead of
STT. 
Particle identification with a ring-imaging Cherenkov
(RICH) counter realized by the detection of internally reflected
Cherenkov light (DIRC) detector will follows. The forward region will
be covered by two sets of mini drift chambers (MDC) and another
Cherenkov detector, either aerogel RICH or a flat DIRC. The inner
detectors are surrounded by an electromagnetic calorimeter
(EMC). 
It is also planned the use of time-of-flight (TOF) counters
for identification of the particles with momenta lower then 5GeV/c.
Outside of the solenoid scintillating bars for muon
identification (MUO) will be mounted. For the more forward directed
particles tracking and possibly a time of flight start signal will be
obtained using the MVD and MDCs.

The current design of the forward spectrometer (FS) includes a 1 m 
gap dipole and tracking detectors like MDC and straw tube
trackers. Photons will be detected by a shashlyk-tipe calorimeter
consisting of lead-scintillators sandwiches (EMC). 
Particle identification via a ring-imaging Cherenkov (RICH)
and time-of-flight counters is previewed.
Other neutrals and
charged particles with momenta close to the beam momentum will be
detected in the hadron calorimeter and muon counters (MUO).

\section{Motivation}
\label{motivazioni}

\begin{figure}[h!] 
\begin {minipage}{0.45\textwidth}
\begin{center}
\includegraphics[width=0.7\textwidth,keepaspectratio]{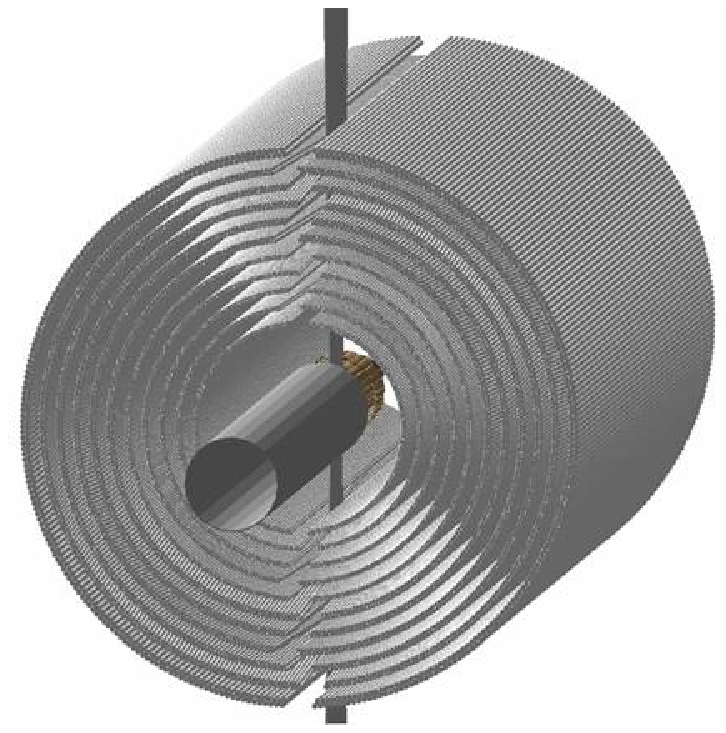}
\caption{\textit{View of Straw Tube Trackers (STT) with 11 double layers.}}
\label{strawtubetracker}
\end{center}
\end{minipage}
\hfill
\begin{minipage}{0.45\textwidth}
\begin{center}
\hspace*{-0.5cm}
\includegraphics[width=\textwidth,keepaspectratio]{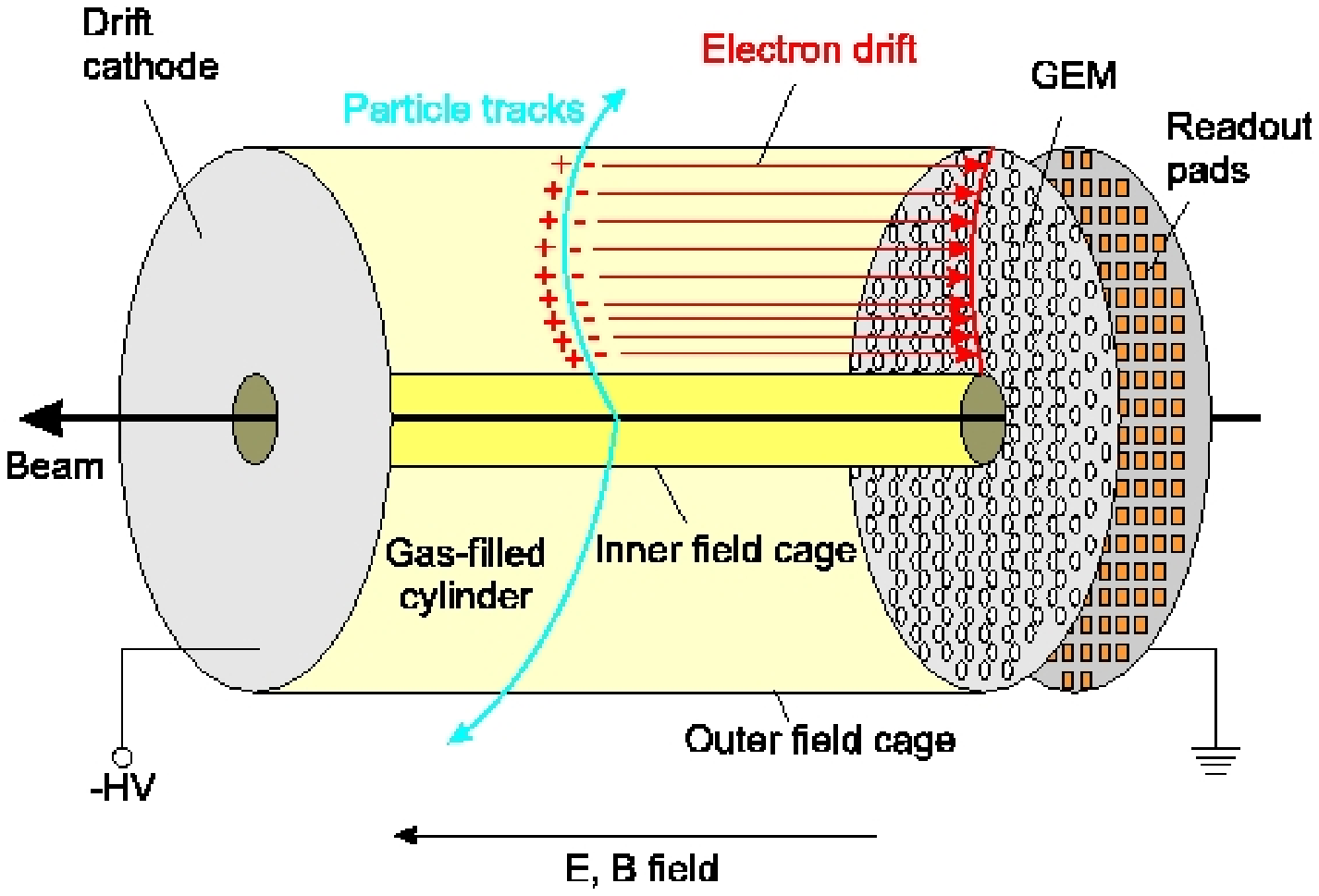}
\caption{\textit{Schematic view of a GEM based Time Projection Chamber
    (TPC).}}
\label{timeprojectionchamber}
\end{center}
\end{minipage}
\end{figure}

To check deeply such an experimental apparatus and to test the
different tracking options the best solution is to
take into account a known channel. The channel $\bar{p}p \rightarrow
\Lambda\bar{\Lambda}$ was intensively investigated by the experiment
PS185 
\cite{PS185web,PS185LOI,PS185-1,PS185-2,PS185-3}
and can offer a good comparison for our collaboration. 
In particular, it is important to compare the behaviour of two
tracking options: the straw tubes tracker (STT) and the time
projection chambers (TPC), shown in Fig. \ref{strawtubetracker} and 
Fig. \ref{timeprojectionchamber}, respectively. 

As main advantages, an STT can offer a robust mechanical stability, an
high tracking efficiency due to minimal dead zones of the tubes and
especially a high rate capability. The main disadvantage of STT is
the amount of dead material that can cause additional multiple
scattering. The TPC tracker can offer high granularity and better
particle identification at low momenta in a continuous operation
mode. On the other hand a TPC tracker is a slower detector due to the
long drift time needed; this can create some problems at high rates,
where tracks of several events will overlay and we have to disentangle
the tracks of each event from other events.



The results of this comparison will be discussed in the following 
paragraphs.
\\

With the proposed detector it will be possible to perform the determination
of spin observables of the reaction $\bar{p}p \rightarrow
\Lambda\bar{\Lambda}$. Taking advantage of the self-analyzing power of
the parity-violating weak decay
$\Lambda \rightarrow p\pi^-$ ($\bar{\Lambda} \rightarrow \bar{p}\pi^+$),
the spin correlation between produced hyperons \cite{paperBelostot}. 
This provides a check of the
factorization approach and of the hadronization process.
When the HESR facility will reach the full design luminosity, it will
also be possible to address other physics topics like $CP$ violation 
in the strangeness sector.

\section{Event generation and simulation}
\label{sec:sim:res:llbar:intro}

The reaction $\bar{p}p \rightarrow \Lambda\bar{\Lambda}$ 
was simulated at beam momenta of 4.0,6.0,7.7 GeV/c.
Two setup scenarios were 
investigated: one setup used the straw tube trackers (STT) and the other 
one used the time projection chambers (TPC).
These simulations were done without using the TOF in the
target spectrometer and TOF and RICH in the forward spectrometer,
presently not implemented, with
the consequence that particle identification for certain momentum and
angular ranges was not possible.
10000 events were generated. 
For simplicity and in order to fully cover the detector acceptance,
the $\Lambda$ and the $\bar{\Lambda}$ particles
were generated isotropically in the CM frame, without any 
decay asymmetry.

The decays $\Lambda \rightarrow p \pi^-$ and $\bar{\Lambda} \rightarrow 
\bar{p} \pi^+$ were studied.
Possible background arises from wrong particle ID assumption of $p-\pi^+$ 
and $\bar{p}-\pi^-$ respectively. 
The PID information was used for proton and pion identification.

At the beam momenta considered, $\Lambda$ and $\bar{\Lambda}$ were
emitted in the laboratory frame at angles between 0 and 35 degrees 
for the lowest beam momentum 
and between 0 and 55 degree for the higher momenta.  

\section{Event Selection}
\label{sec:sim:res:llbar:sel}

Some cuts on the decay vertex and on the $\Lambda$($\bar{\Lambda}$) 
invariant mass have
been applied for the event selection. 
Cuts are introduced on the $\chi^2$ value of the refitted decay vertex
($\chi^2<$ 5.0)
and on the $z$ coordinate of the decay vertex, which is taken  between 
-2.0 mm and
350.0 mm. Both these selections lead to $\sim$5\% event reduction. The
angle between $\Lambda$($\bar{\Lambda}$) momentum and the direction of
the vector between primary and secondary vertex was also investigated:
it was decided to cut angular discrepancies higher than 1$^\circ$ and due to
this we have an event loss of $\sim$5\%. It was also chosen to cut the
tails of the $\Lambda$($\bar{\Lambda}$) invariant mass distribution
accepting a deviation of 0.003 GeV/c$^2$ from the $\Lambda$ mass value of
1.1157 GeV/c$^2$; this leads to an event reduction of $\sim$1\%.

\section{Results}
\label{sec:sim:res:llbar:results}

The obtained results will reflect the actual status of the
PANDA simulation software.

\begin{figure}[h!] 
\begin {minipage}{0.45\textwidth}
\begin{center}
\includegraphics[width=\textwidth,keepaspectratio]{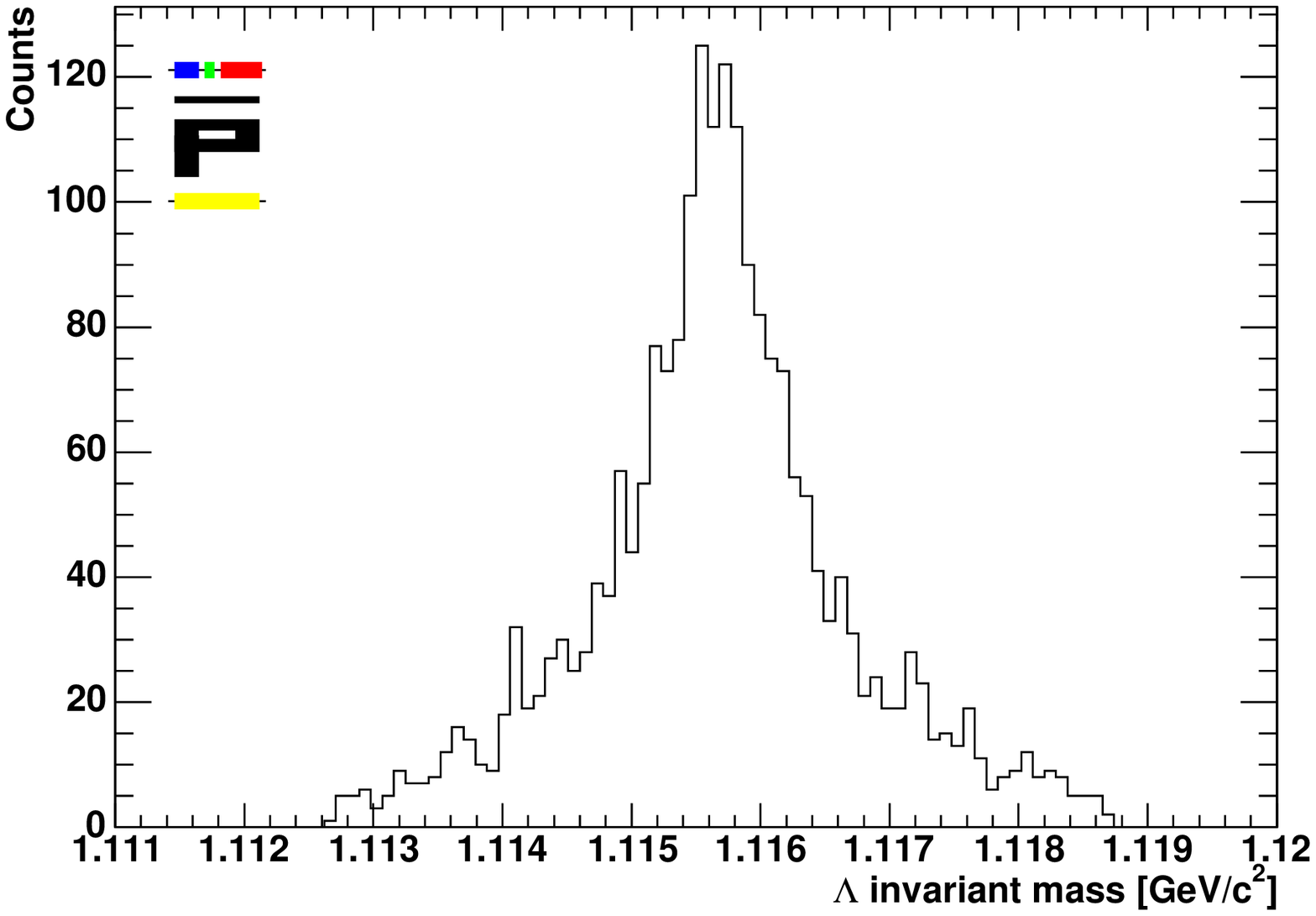}
\caption{\textit{$\Lambda$ invariant mass spectrum at the beam momentum 
4GeV/c$^{2}$ using STT. Mass resolution: 0.71$\pm$0.02 MeV/c$^2$.}}
\label{invmass4stt}
\end{center}
\end{minipage}
\hfill
\begin{minipage}{0.45\textwidth}
\begin{center}
\hspace*{-0.5cm}
\includegraphics[width=\textwidth,keepaspectratio]{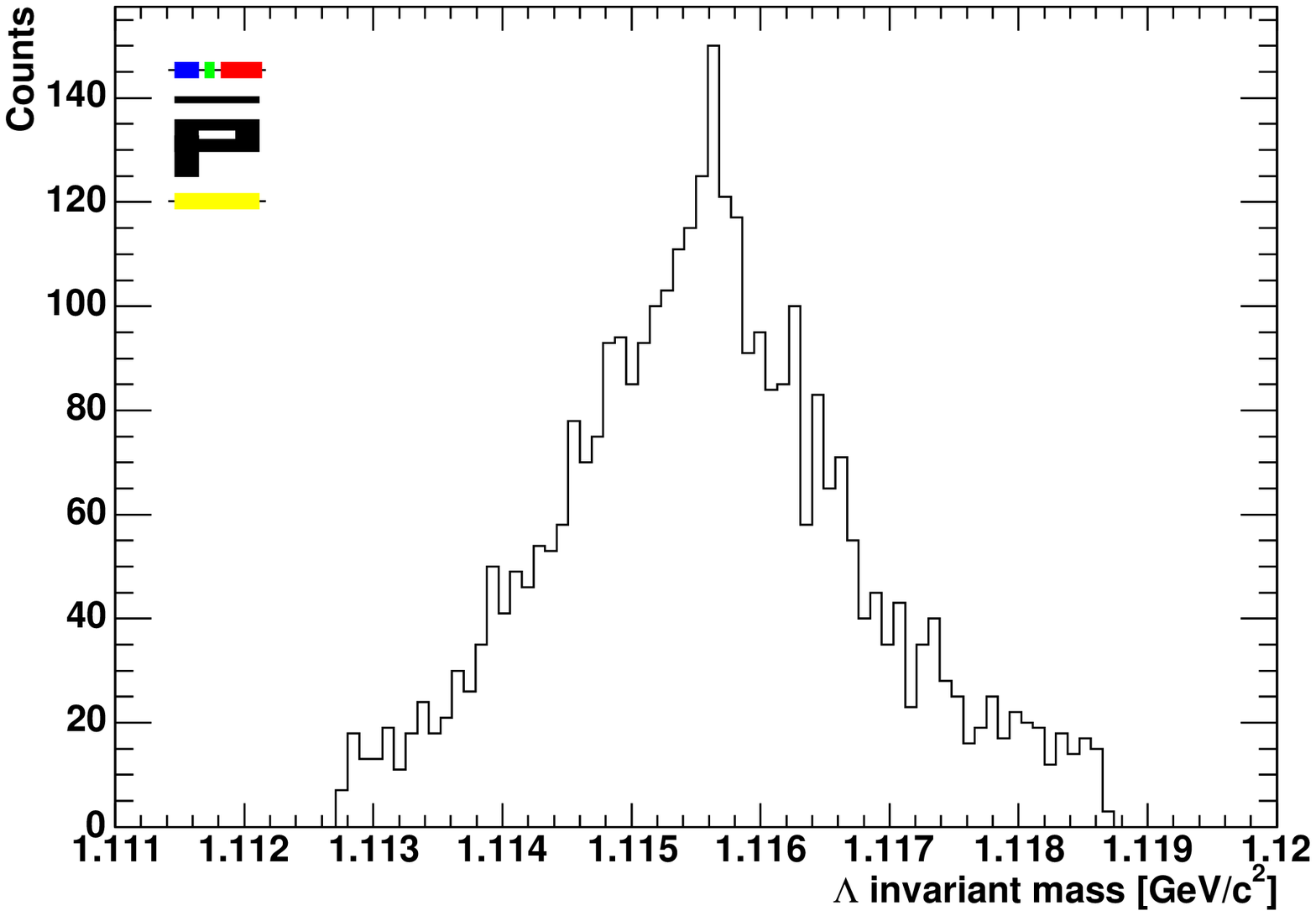}
\caption{\textit{$\Lambda$ invariant mass spectrum at the beam momentum 
4GeV/c$^{2}$ using TPC. Mass resolution: 1.10$\pm$0.02 MeV/c$^2$.}}
\label{invmass4tpc}
\end{center}
\end{minipage}
\end{figure}

The $\Lambda$ invariant mass for the beam momentum of 4.0 GeV/c,
reconstructed from 10000 generated events,
is presented in Fig. \ref{invmass4stt}, obtained with the STT setup, 
and Fig. \ref{invmass4tpc} obtained with the TPC setup, respectively.

Proceeding from the $\Lambda$ momenta distribution it is clear that the TPC
setup allows particle ID even for the slowest tracks. 
In fact it is possible to
reconstruct $\Lambda$ with a momentum between 1 and 4 GeV/c using the
STT setup, while the TPC can reconstruct them starting from 0.5 GeV/c.

With the STT setup, currently the reconstruction efficiency is only 
of the order 
of 25\%, which might be due to a loss of low momentum pions in the 
reconstruction procedure. Clearly, further optimization needs to be done.
Using the STT setup, the reconstruction efficiency of $\bar{\Lambda}$
is lower by 10\% as compared to $\Lambda$ reconstruction. This effect is 
not present when using the TPC. However, the overall reconstruction 
efficiency using the TPC reaches only 35\%. Mass resolutions of 0.71$\pm$0.02 
MeV/c$^2$ for STT and 1.10$\pm$0.02 MeV/c$^2$ for TPC trackers are obtained,
independent of the beam momentum. The tracking options are actually under
study to get a better improvement also for the mass resolutions.

\begin{figure}[h!] 
\begin {minipage}{0.45\textwidth}
\begin{center}
\includegraphics[width=\textwidth,keepaspectratio]{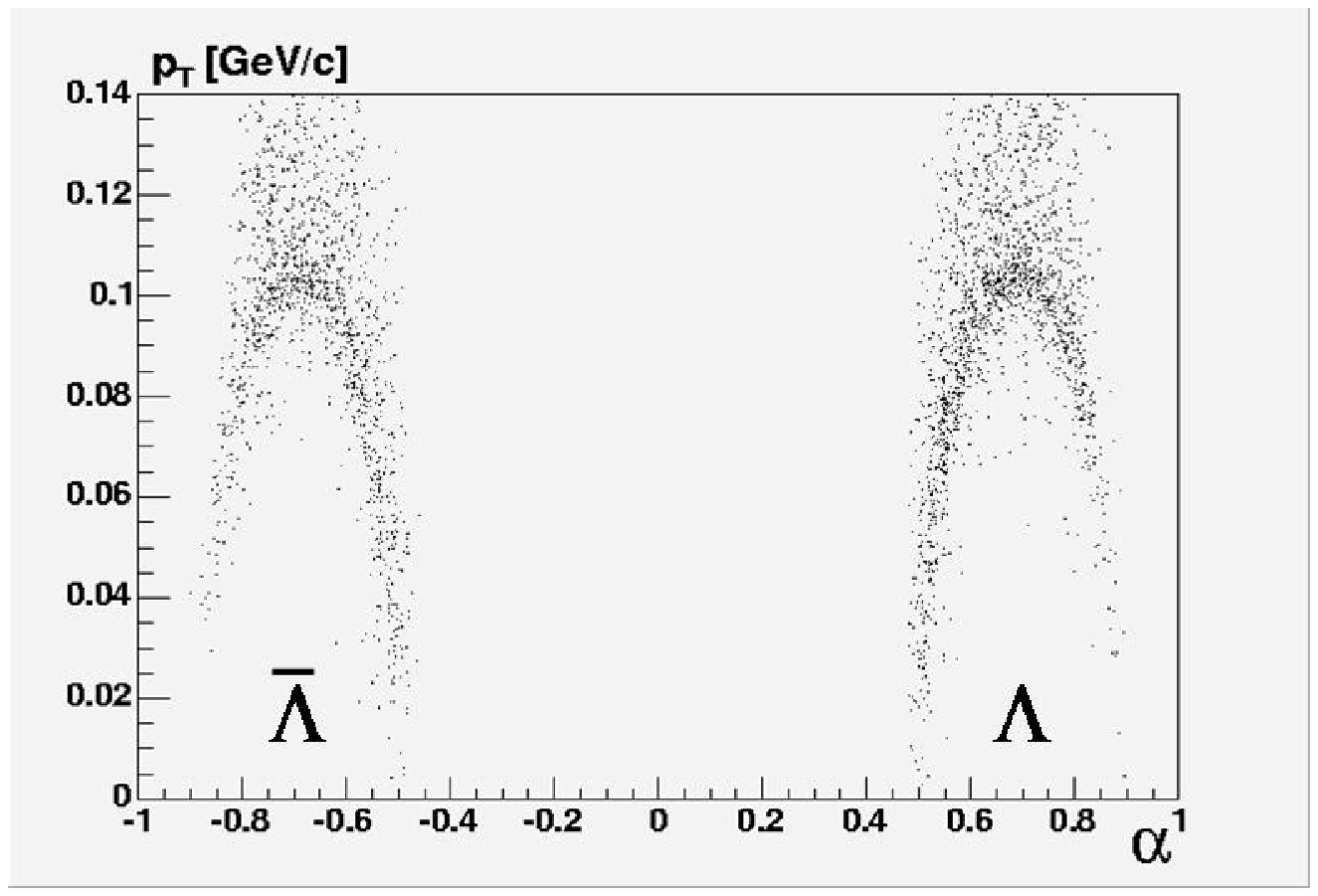}
\caption{\textit{Armenteros plot obtained with STT.}}
\label{armenterosstt}
\end{center}
\end{minipage}
\hfill
\begin{minipage}{0.45\textwidth}
\begin{center}
\hspace*{-0.5cm}
\includegraphics[width=\textwidth,keepaspectratio]{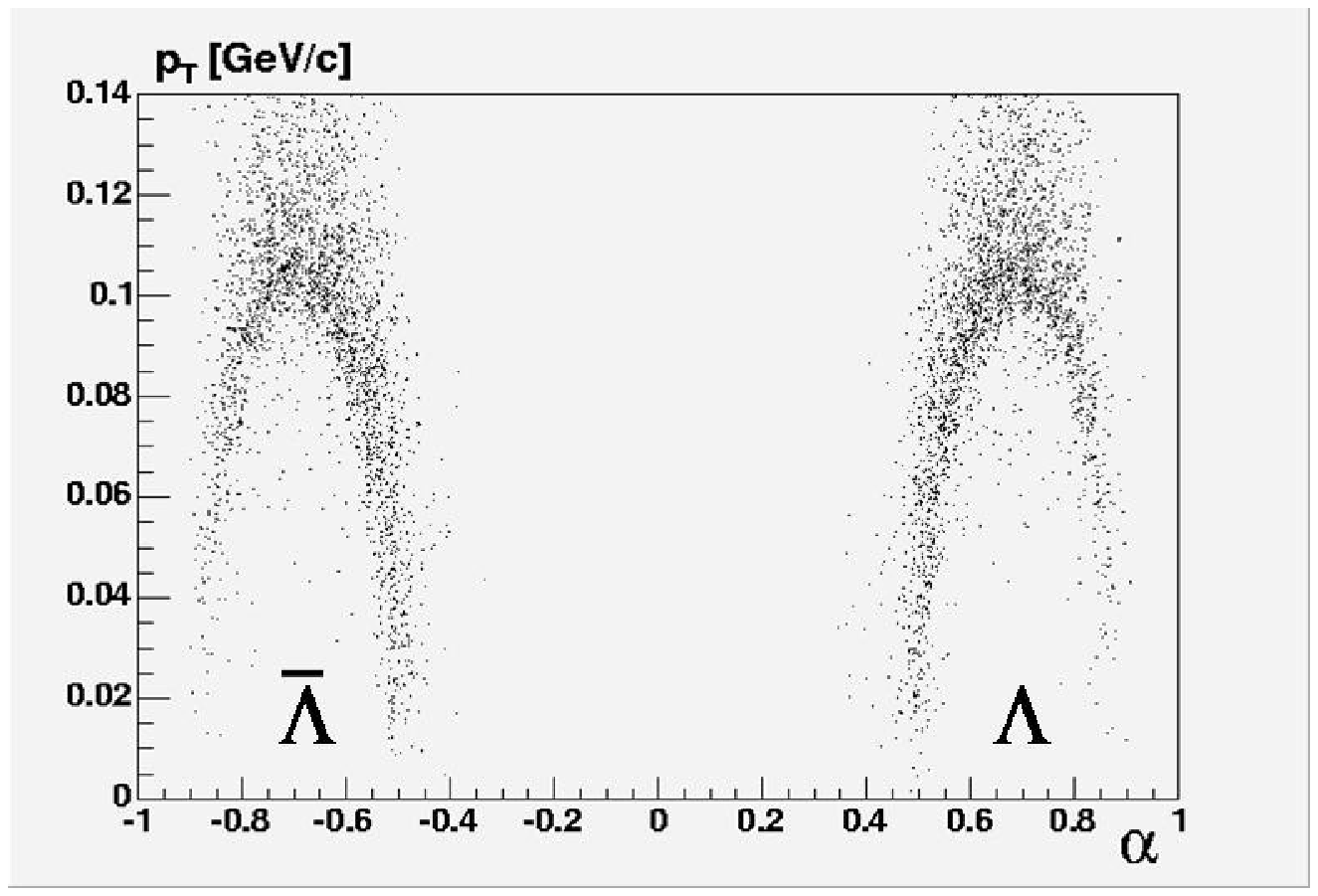}
\caption{\textit{Armenteros plot obtained with TPC.}}
\label{armenterostpc}
\end{center}
\end{minipage}
\end{figure}

These peculiarities are also visible in an Armenteros plot. In this
plot, the 
transverse momentum of the decay products in the $\Lambda$ rest frame 
is plotted as a function of the asymmetry $\alpha$, where $\alpha =
\frac{p_{L+}-p_{L-}}{p_{L+}+p_{L-}}$ and $p_{L+(-)}$ is the
longitudinal momentum of the positive (negative) particle produced in
the decay. In Fig. \ref{armenterosstt} and Fig. \ref{armenterostpc} the 
Armenteros plot obtained for both setups are shown.

For the $\Lambda\bar\Lambda$ pair reconstruction with the STT setup it 
is possible to 
reach an efficiency of 6\%. 
Using the TPC setup, instead, the reached efficiency is around 25\%.
The efficiency clearly need to be improved for both the tracking options.
The missing momentum, the missing energy and the missing mass
were studied for the different beam momenta.
In Table \ref{tab:resolutionLLbar} the resolutions obtained with a beam 
momentum of 4.0 GeV/c for both setups are shown.
It should be noticed that the missing momentum resolution is higher 
in transverse direction as compared to longitudinal direction.
Comparing the STT and the TPC trackers, the STT 
setup offers a better resolution, but a lower reconstruction efficiency.
With both setups, the resolution of the longitudinal missing momentum is 
worst at the highest beam momenta. 
Every value of this table should be centred in 0.
This indicates the presence of some discrepancies in our simulation or
reconstruction routines that needs to be investigated.

\begin{table}[h]
\begin{center}
\begin{tabular}{|l|c|c|}
\hline
                                     & STT $mean\pm\sigma$   & TPC $mean\pm\sigma$ \\\hline
Missing Energy [MeV]                 & 19.98$\pm$0.88 & 39.54$\pm$0.88 \\\hline
Missing Mass [MeV/c$^2$]             & 46.43$\pm$2.12 & 80.48$\pm$2.95 \\\hline
Missing Transverse &&\\
Momentum [MeV/c]                     & 9.26$\pm$0.33  & 13.70$\pm$0.20 \\\hline
Missing Longitudinal &&\\
Momentum [MeV/c]                     & 20.28$\pm$0.79 & 35.25$\pm$1.05 \\
\hline
\end{tabular}
\end{center}
\caption{Resolution of missing energy, missing mass and missing momentum 
for a $\Lambda$ $\bar{\Lambda}$ couple at a beam momentum of 4.0
GeV/c. The mean value $\pm$ the sigma of the gaussian distribution 
obtained is presented.}
\label{tab:resolutionLLbar}
\end{table}  
 
The different beam momenta studied (6.0 and 7.7 GeV/c) give comparable 
results with the one shown here for a beam momentum of 4 GeV/c.

\section{Summary}
\label{summary}

The reconstruction of $\Lambda$ $\bar{\Lambda}$ pairs is possible with
the PANDA spectrometer.

The use of a known channel ($\Lambda\bar{\Lambda}$) gives the
possibility to perform the study of the PANDA detector acceptance,
resolution and background suppression. In this way, it is also
possible to obtain a good comparison between different tracking
options, that in the case of my work were Straw Tube Tracker (STT) and
Time Projection Chamber (TPC).

These first simulation results pointed out some software problems that
need to be fixed in the future. 
It is also important to add the PID detectors in the target 
and the forward spectrometer (TOF and RICH) in the simulation geometry to
have more improvement.
Background studies based on the Dual Parton Model \cite{DPM} are in progress.
Furthermore, we plan to study the decay asymmetry reconstruction
of the $\Lambda$ and the $\bar\Lambda$ particles.

\section{Aknowledgements}
\label{aknowledgements}

This work was supported in part by DFG, EU, GSI and BMBF.

\end{document}